\documentstyle[twoside,fleqn,espcrc2,epsf]{article}

\title{Heavy Hybrids From NRQCD}
\author{Presented by S.~Collins, UKQCD Collaboration\address{Dept. of Physics and Astronomy, Glasgow University, Glasgow, G12 8QQ, Scotland}
\thanks{In collaboration with C.~Davies, Glasgow Univ.. Results provided by G.~Bali, Southampton Univ. are also presented.}}
       
\begin{document}

\begin{abstract}
We present results for the exotic $0^{+-}$ and $1^{-+}$ $b\bar{b}g$
hybrids at $\beta=6.0$. A leading order NRQCD action is used
to generate the heavy quark propagators. The optimisation of operators
is discussed and we compare our results to the predictions of recent
potential calculations.
\end{abstract}
\maketitle
\section{INTRODUCTION}

In hybrid mesons the gluon string between the $q\bar{q}$ pair is
excited and contributes to the quantum numbers of the particle. These
mesons are predictions of QCD which are not yet confirmed in
experiment. More difficult to discriminate from quark model states
than glueballs, it is particularly important to provide accurate
predictions from lattice studies.

Most experimental candidates are for hybrids containing light quarks;
however, the best system for a hybrid search may be $c\bar{c}$ or
$b\bar{b}$, where there is a large gap between the lowest state and
the $D\bar{D}$ and $B\bar{B}$ thresholds respectively.  Theoretically,
heavy hybrids are easier to study; on the lattice, NRQCD can be used
to simulate heavy quarks and the results can be compared with those
from hybrid potentials.


Here, we present results for $b\bar{b}g$ hybrids.
The lattice hybrid operators are constructed from staples
of gauge links, $\sqcup{-}\sqcap$, in combinations corresponding to the
$T_1^{+-}$ and $T_1^{-+}$ representations of the cubic symmetry goup,
as described in Lacock et al.~\cite{lacock}. In early potential
studies~\cite{perant_n_chris} it was found that this gluon excitation,
corresponding to one unit of angular momentum about the $q\bar{q}$
axis, gives rise to the lowest hybrid meson states. Combined with 
$q\Gamma\bar{q}$ in a spin singlet~($A_1^{-+}$) and triplet~($T_1^{--}$)
the hybrid meson states
\begin{eqnarray}
T_1^{+-}: & &  1^{--},O^{-+},{\bf 1^{-+}},2^{-+}\nonumber\\
T_1^{-+}: & &  1^{++},{\bf O^{+-}},1^{+-},{\bf 2^{+-}}\nonumber
\end{eqnarray}
are accessible, where the exotic states are shown in bold. In the
static limit these states are all degenerate; introduction of quark
motion, using the lowest order NRQCD action, will create a splitting
between the sets of states arising from the $T_1^{+-}$ and $T_1^{-+}$
gauge paths. Spin-spin and spin-orbit interactions are needed in order
to obtain splittings between the individual states. These spin
splittings, eg between the $O^{-+}$, $1^{-+}$ and $2^{-+}$, will be
ordered in the same way as for the non-hybrid $^3P$ states and are likely to be
of order $10$~MeV. Since the statistical errors in our
calculation are still at approximately $200$~MeV we ignore the fine structure
and use the lowest order NRQCD action including only the
kinetic energy term.
In this case there is no mixing between the hybrid and normal
$b\bar{b}$ states. 

The correlators corresponding to the hybrid operators were computed on
500 quenched $16^3\times 48$ configurations at $\beta=6.0$. A bare $b$
quark mass, from $b\bar{b}$ spectroscopy, of $1.71a^{-1}$ was used in the
NRQCD action. The hybrid signal is dominated by noise after only a few
timeslices, and multiple spatial and temporal sources are required. It
is essential to optimise the operators in terms of the $q\bar{q}$
separation, $R$, depth of the staple, $D$, and the coefficient, $c$,
and level, $n$, of the fuzzing of the gauge links. We use the
results of a study of the hybrid potential~\cite{bali} on the same
configurations to choose the optimal values of these parameters.

\section{THE HYBRID POTENTIAL}

The hybrid potential corresponding to the $T_1$ representation, $E_u$, was
calculated in the standard way using operators of the form $\sqcup{-}\sqcap$.
In order to find the optimal operator, the depth $D$ was varied at each $R$
and the maximal overlap of the operator with the ground state,
$A_{g.s.}$, was sought. Table~\ref{overlaps} presents the results
as a function of $D$ for the example of $R=6$. $A_{g.s.}$ is not
very sensitive to $D$ as long as $D$ is small.  A fairly uniform optimal
overlap was found for a wide range of $R$, where we investigated up to $R=9$.

\begin{table}
\caption{}
\begin{tabular}{|c|c|c|c|c|}\hline
$R\times D$ & $6\times 1$ &  $6 \times 2$ & $6 \times  3$ & $6 \times 6$ \\\hline
$A_{g.s.}$ & 0.90(2) & 0.92(2) & 0.92(2) & 0.39(8) \\\hline
\end{tabular}
\label{overlaps}
\end{table}

Fuzzing the gauge links seemed to act as a smoothing function on the
hybrid source and it was found sufficient to optimise the value of the
spatial plaquette with respect to $c$ and $n$ in order to obtain their
optimal values: roughly $c=2.3$ and $n=25$.

\begin{figure}
\vskip -0.5cm
\epsfxsize=7cm\epsfbox{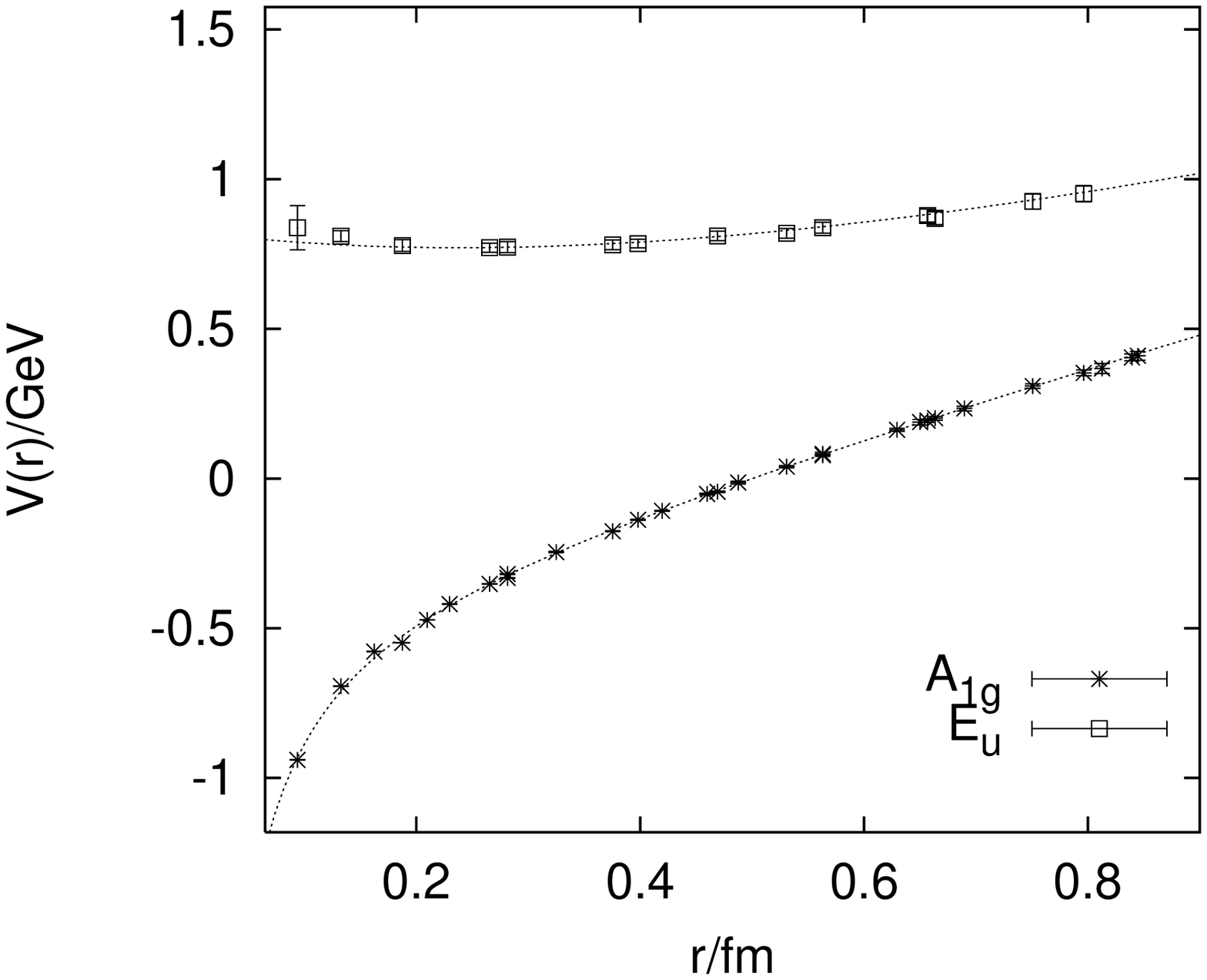}
\vskip-1cm
\caption{\vskip -0.6cm}
\label{wave}
\end{figure}

\begin{figure}
\vskip -0.5cm
\epsfxsize=7cm\epsfbox{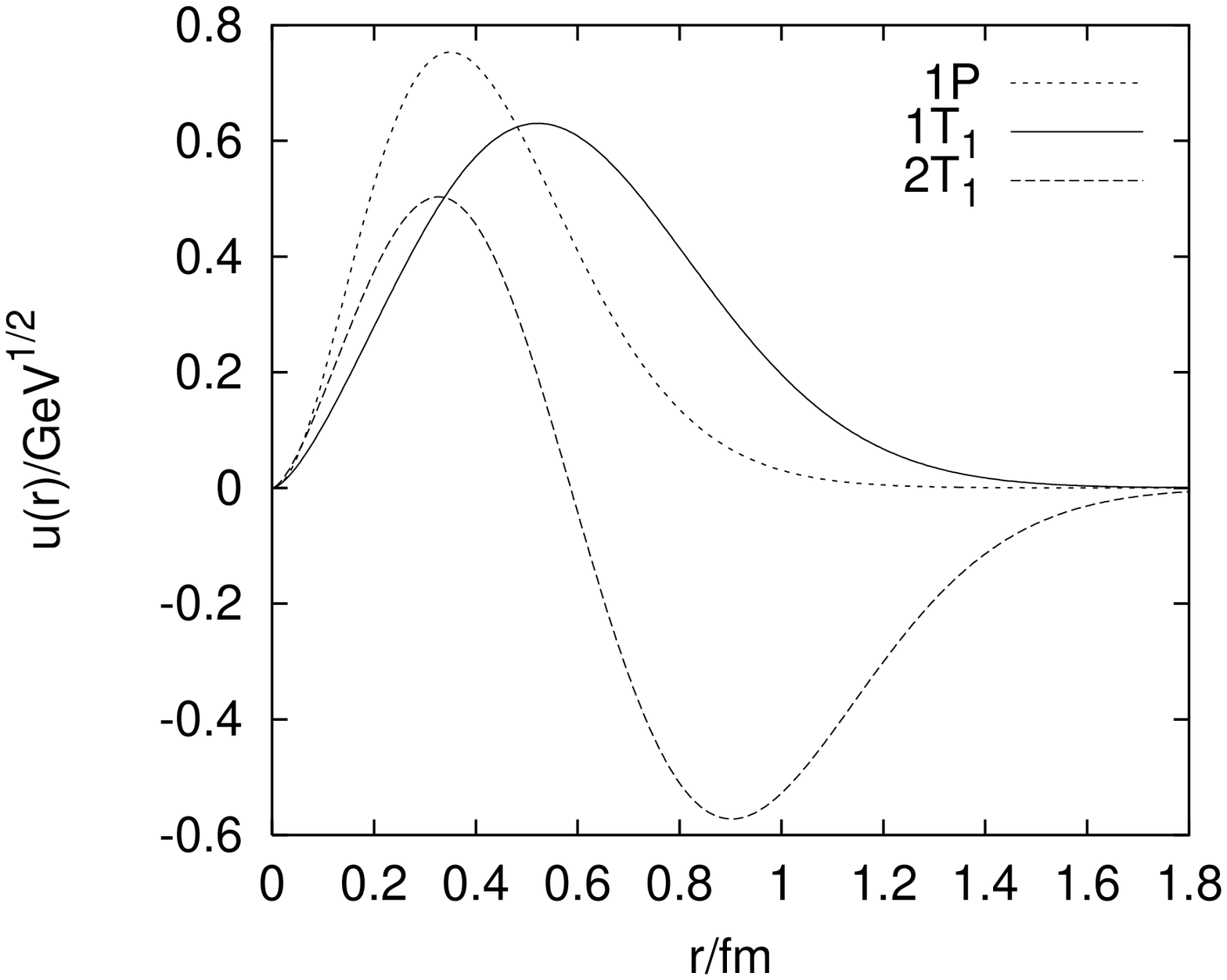}
\vskip-1cm
\caption{\vskip -0.6cm}
\label{hybwave}
\end{figure}

Using these optimal parameters for the gauge path, the hybrid
potential shown in figure~\ref{wave} was obtained; the ground state
potential, $A_{1g}$, is also shown for comparison. The most striking
feature of the potential is its flatness which suggests broad hybrid
wavefunctions and densely packed states. To extract this information
the potential was fitted to a functional form suggested by string
models:
\begin{equation}
V = V_{g.s.} + \left(1 - e^{-cR}\right)\pi/R,
\end{equation}
with parameter $c$.
The hybrid energy levels, $E_{nl}$, and wavefunctions, $u_{nl}$, where
$\psi = u_{nl} Y^{nl}_m/R$, can then be extracted by solving the
Schr\"odinger equation in the continuum.
The lowest lying hybrid levels are given in table~\ref{potlevel} in GeV and
the corresponding radial wavefunctions are shown in figure~\ref{hybwave}. The
scale was set using the average of the lattice spacings from the $b\bar{b}$
$P-S$ and $2S-1S$ splittings. Note that the first hybrid state is a $P$-state.
The $1P$ $b\bar{b}$ wavefunction is shown for comparison in the figure.
%
\begin{table}
\caption{}
\begin{tabular}{|c|c|c|c|}\hline
n &  1 & 2 & 3 \\\hline
$l=1$ & 10.87 & 11.13 & 11.39\\\hline
$l=2$& 11.02 & 11.29 & 11.55\\\hline
\end{tabular}
\label{potlevel}
\end{table}

The wavefunctions are broad, as expected. For a $L\sim 1.8$~fm lattice
we can expect large finite volume errors for a ground state hybrid
meson with $r_{r.m.s.}\sim 0.28$~fm. Figure~\ref{hybwave} suggests a
lattice hybrid operator with a $q\bar{q}$ separation of $0.5$~fm or
$\sim 6$ lattice units at $\beta=6.0$ would minimise the overlap with
the first excited state since the corresponding wavefunction has a
node at this point. However, a more reliable method to extract the
ground state energy is to perform multiple exponential fits to a
matrix of correlation functions obtained from different $q\bar{q}$
separations, where the different values of $R$ give rise to very
different overlaps with the hybrid excited states in the off-diagonal
elements of the matrix. For example, from figure~\ref{hybwave} we see a
suitable $2\times2$ matrix can be formed from $R=3$ and $R=7$.
%

%


\section{HYBRIDS WITH PROPAGATING QUARKS}
\begin{figure}
\epsfxsize=5.9cm\epsfbox{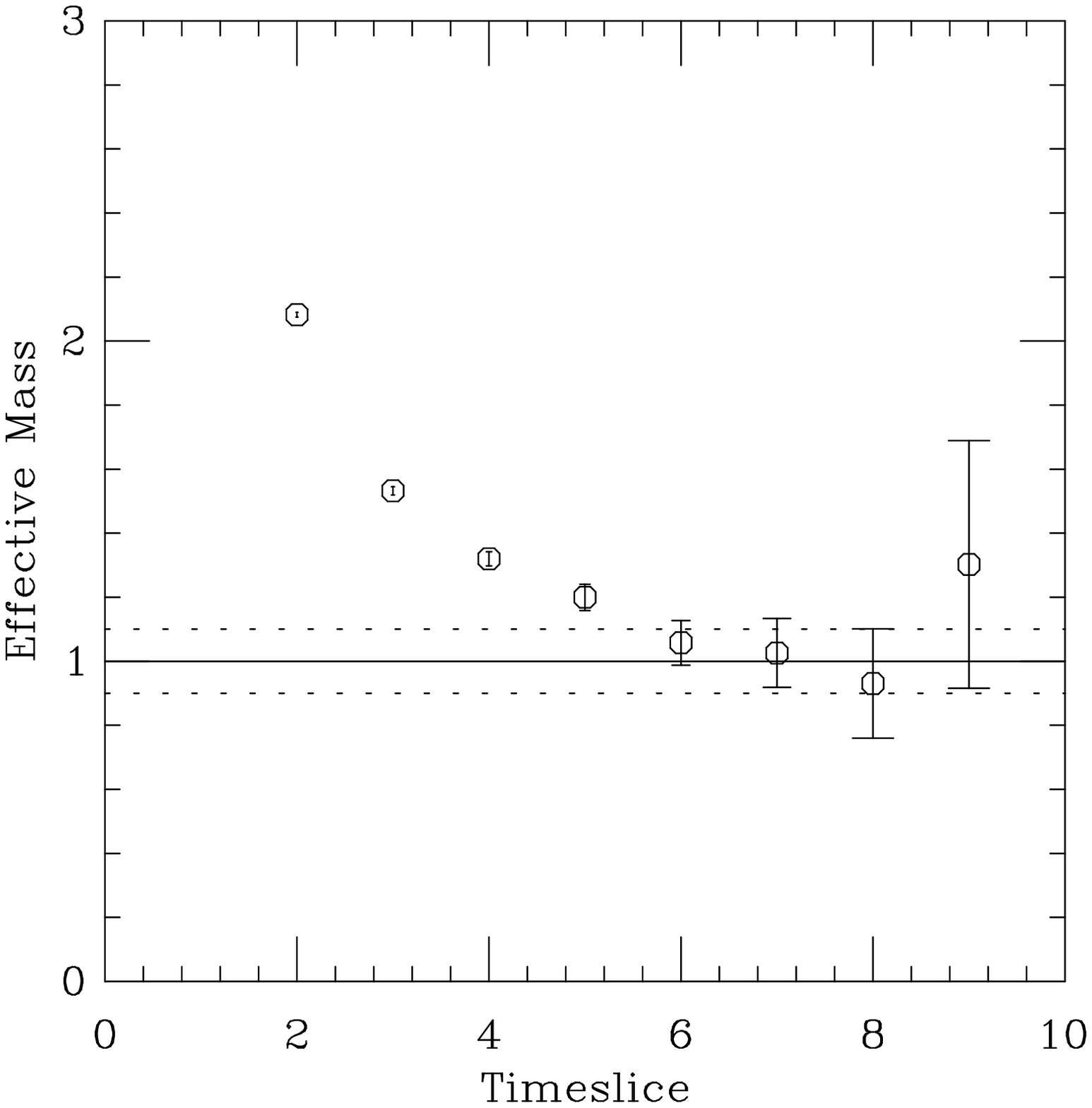}
\vskip -1cm
\caption{\vskip -0.6cm}
\label{eff}
\end{figure}

Considering the potential results we performed an initial
calculation of the $T_1^{+-}$ and $T_1^{-+}$ hybrids using a staple
size of $R=3$ and $D=2$ at both the source and the sink. Figure~\ref{eff}
presents the results for the effective mass of the $T_1^{+-}$ hybrid
obtained from $10,000$ sources. There is a large contribution
from excited states for this operator. However, a plateau seems to
begin from timeslice $6$ or $7$. One-exponential fits in the region
$6{-}10$ are in agreement with two-exponential fits with $t_{min}=3{-}5$.
Considering all these fits, we find the ground state
simulation energy to be consistent with $1.0(1)$, as indicated in the
figure. This corresponds to a splitting with the $^3S_1$ $b\bar{b}$
state of $0.55(10)$ in lattice units. We also find a splitting of
$0.1-0.3$ with the $T_1^{-+}$ states, which confirms that the $1^{-+}$
is the lowest exotic hybrid.

\begin{figure}
\epsfxsize=7cm\epsfbox{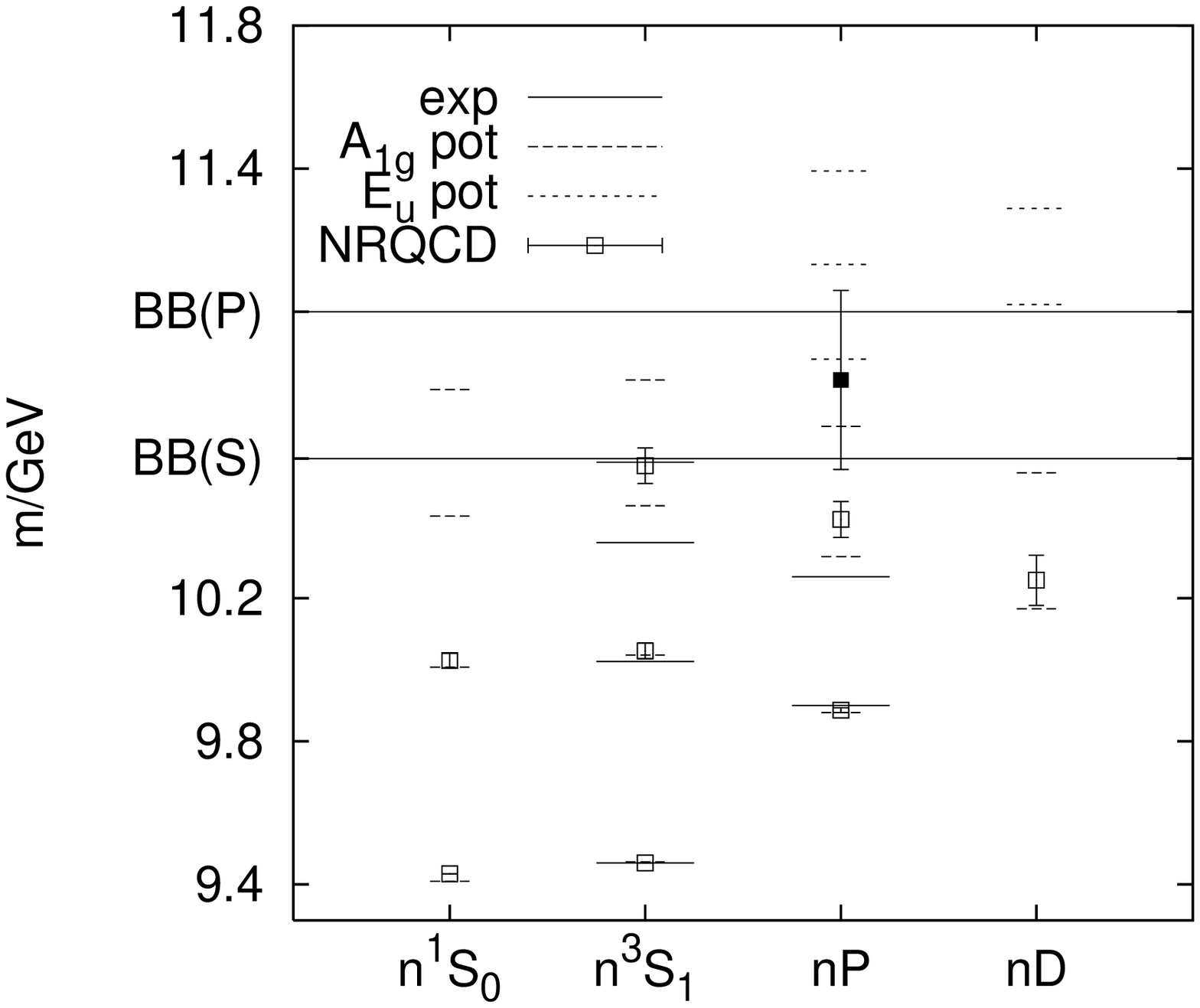}
\vskip -1cm
\caption{\vskip -0.6cm}
\label{expt}
\end{figure}

Converting our results into physical units we obtain $11.5(5)$~GeV for
the $T_1^{+-}$ hybrid state. This is compared with the lower lying
$b\bar{b}$ spectrum obtained from an $O(Mv^4)$ NRQCD
action~\cite{chris} and also from a potential study, accurate to the
same order in $v$~\cite{balipot} in figure~\ref{expt}. Our prediction
lies $1\sigma$ above the threshold for $B\bar{B}$ production but
$1\sigma$ below that for $BB^{**}$, which is predicted by flux tube
models to be the dominant hybrid decay mode. A study by Manke et
al..~\cite{thomas} using different operators on the same
configurations finds consistent results with the $T_1^{+-}$ hybrid
lying $1\sigma$ higher.  Our results are also consistent with the
predictions from the potential.

The position of the hybrid state with respect to the two thresholds is
not settled. An improved determination of the hybrid state is needed
and we propose to do this by computing a matrix of correlators between
different $q\bar{q}$ separations. At the moment the large statistical
errors dominate the uncertainty in the prediction. However, the
central value may be shifted significantly by finite volume effects
and to a lesser extent quenching and the systematic errors associated
with NRQCD.

%
\section*{ACKNOWLEDGEMENTS}
The computations were performed on the J90 at EPCC in
Edinburgh as part of HPCI under EPSERC grant
no. GR/K55745.


\begin{thebibliography}{99}
\bibitem{lacock} P.~Lacock et al., Phys.~Rev.~D55 (1997) 1548.
\bibitem{perant_n_chris} C.~Michael et al., Nucl. Phys. B347 (1990) 854.
\bibitem{bali} G.~Bali, in preparation.
\bibitem{chris} C.~Davies et al., Phys.~Rev.~D50 (1993) 6963.
\bibitem{balipot} G.~Bali et al., Phys.~Rev.~D56 (1997) 2566.
\bibitem{thomas} T.~Manke, this publication.
\end{thebibliography}
\end{document}